\documentclass[prl,superscriptaddress,twocolumn,amsmath,amssymb,keywords,floatfix,showpacs]{revtex4}
\usepackage{bm}
\usepackage{epsfig}
\usepackage{graphicx}
\usepackage{color}
\usepackage{nicefrac}

\begin{document}

\title{Muonic vacuum polarization correction to the bound-electron $g$-factor}

\date{\today}

\author{N.~A.~Belov}
\affiliation{Max Planck Institute for Nuclear Physics, Saupfercheckweg 1, 69117 Heidelberg, Germany}
\author{B.~Sikora}
\affiliation{Max Planck Institute for Nuclear Physics, Saupfercheckweg 1, 69117 Heidelberg, Germany}
\author{R.~Weis}
\affiliation{Max Planck Institute for Nuclear Physics, Saupfercheckweg 1, 69117 Heidelberg, Germany}
\author{V.~A.~Yerokhin}
\affiliation{Max Planck Institute for Nuclear Physics, Saupfercheckweg 1, 69117 Heidelberg, Germany}
\affiliation{Center for Advanced Studies, Peter the Great St.~Petersburg Polytechnic University, 195251 St.~Petersburg, Russia}
\author{S.~Sturm}
\affiliation{Max Planck Institute for Nuclear Physics, Saupfercheckweg 1, 69117 Heidelberg, Germany}
\author{K.~Blaum}
\affiliation{Max Planck Institute for Nuclear Physics, Saupfercheckweg 1, 69117 Heidelberg, Germany}
\author{C.~H.~Keitel}
\affiliation{Max Planck Institute for Nuclear Physics, Saupfercheckweg 1, 69117 Heidelberg, Germany}
\author{Z.~Harman}
\affiliation{Max Planck Institute for Nuclear Physics, Saupfercheckweg 1, 69117 Heidelberg, Germany}


\begin{abstract}

The muonic vacuum polarization contribution to the $g$-factor of the electron bound in a nuclear potential
is investigated theoretically. The electric as well as the magnetic loop contributions are evaluated.
We found these muonic effects to be observable in planned trapped-ion experiments with light and medium-heavy highly charged ions.
The enhancement due to the strong Coulomb field boosts these contributions much above the corresponding terms in the free-electron
$g$-factor. Due to their magnitude, muonic vacuum polarization terms are also significant in planned determinations of the fine-structure
constant from the bound-electron $g$-factor.

\end{abstract}

\maketitle

\begin{figure}[t]
   \includegraphics[width=1.0 \columnwidth]{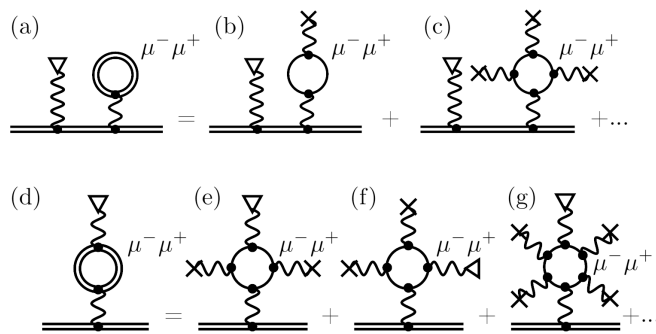}
   \caption{Feynman diagrams of the one-loop muonic vacuum polarization terms: (a--c) electric and (d--g) magnetic loop corrections.
   Single loop lines denote free muon propagators, double loop lines represent Coulomb-Dirac muon propagators, and double straight lines represent
   Coulomb-Dirac electronic wave functions or propagators.
   A wave line terminated by a triangle represents the interaction with the magnetic field, while a wave line terminated by a cross
   denotes the Coulomb interaction with the nucleus.}
   \label{fig:fig1}
\end{figure}

Studies on the bound-electron $g$-factor allowed one of the most stringent tests of strong-field quantum electrodynamics (QED)~\cite{Stu11,Wag13,Stu13}.
From the comparison of theoretical values for hydrogenlike ions with the corresponding experiments,
the most accurate values of the electron mass have been obtained recently~\cite{Stu14,Haf00,Ver04}.
$g$-factor experiments with highly charged ions are anticipated to provide an independent means to determine or even
improve the value of the fine-structure constant $\alpha$~\cite{Sha06,Vol14,Yer16}.
Correspondingly accurate predictions can be achieved by the incorporation of all significant QED effects.
Therefore, the Dirac value of the $g$-factor, apart from the finite nuclear size~\cite{Gla02,Zat12} and mass corrections~\cite{Koh16},
also needs to be extended by the well-known radiative corrections: the self-energy and vacuum polarization (VP) terms (see e.g.~\cite{Bei00full}).
In case of the $g$-factor one can differentiate between electric and magnetic loop VP contributions, depending on whether the nuclear Coulomb or the external 
magnetic field is corrected by a virtual particle loop [see Figs.~\ref{fig:fig1}~(a) and (d)].
Also, QED corrections on the two-loop level have been completely evaluated up to 4th order in the Coulomb field's strength parameter $Z\alpha$~\cite{Pac04,Pac05}, with
$Z$ being the atomic number.

In this Letter we study an extension of the VP corrections, namely, the case when the VP loop is formed by charged leptons heavier than the electron and the
positron. The largest contribution of this type is due to the virtual creation and annihilation of a muon pair.
All calculations have been performed by including the interaction of the bound electron and the dilepton pair with the nucleus to all orders.
As we show, this effect needs to be included in projected studies~\cite{Qui01,Klu08,Stu13AP} with intermediate- and high-$Z$ ions.
Its contribution also needs to be accounted for in planned improvements of the fine-structure constant~\cite{Sha06,Vol14,Stu13AP,Yer16}.
Furthermore, the evaluation of these corrections opens a new way of testing muonic effects in rather compact trapped-ion experiments
and without the need of creation of short-lived muonic atoms~(see e.g.~\cite{Ant13}) or large-scale accelerator or laser
facilities~\cite{Mom87,Mul08}. We show that planned experiments~\cite{Stu13,Qui01} with intermediate- and high-$Z$ ions
can be more sensitive to muonic VP terms than the best current studies of the free-electron $g$-factor~\cite{Han08}.

In the following we discuss separately the evaluation of contributing muonic VP-diagrams: the electric loop and the magnetic loop VP
corrections. Finally, we discuss the significance of these contributions in comparison with each other and
with other known effects contributing to the $g$-factor, such as, e.g., the finite nuclear size, and experimental prospects of observing them.

\textit{Electric-loop vacuum-polarization contributions.} --
The electric-loop VP effects
can be described by radial potentials (see e.g.~\cite{Bei00full}), which correct the pure nuclear potential. These VP potentials,
evaluated separately with one interaction vertex of the free pair with the nucleus, and with three or more vertices, are the Uehling
[see Fig.~\ref{fig:fig1}~(b)] and Wichmann-Kroll (WK) [see Fig.~\ref{fig:fig1}~(c)] potentials, respectively. The Uehling potential
is given by
\begin{eqnarray}
&&V_{\rm Ue}(r)=-\alpha Z\frac{2\alpha}{3\pi}\int_0^\infty \mathrm{d}r'4\pi r'\rho_{\rm nucl}(r')\\
&&\times\int_1^\infty \mathrm{d}t\left(1+\frac{1}{2t^2}\right)\frac{\sqrt{t^2-1}}{t^2}
\frac{e^{-2m|r-r'|t}-e^{-2m(r+r')t}}{4mrt},\nonumber
\end{eqnarray}
where $Z\rho_{\rm nucl}(r)$ is the nuclear charge distribution, and $m$ the mass of the virtual particle in the VP loop (in our case,
the muon mass $m_{\mu}$). The WK potential is given by~\cite{Sof88,Art01}:
\begin{equation}
V_{\rm WK}(r)=-4\pi\alpha\left(\frac{1}{r}\int_0^r\rho(r'){r'}^2\mathrm{d}r'+\int_r^\infty\rho(r')r'\mathrm{d}r'\right),
\end{equation}
with the WK charge density
\begin{eqnarray}
\rho(x)&=&\frac{e}{2\pi^2}\int_0^\infty \mathrm{d}u \label{eqn:WKrho}\\
&&\left(\sum_{\kappa=\pm 1}^{\pm\infty}\right.\left|\kappa\right| {\rm Re}
      \left(\sum_{a=1}^{2}\right.G_\kappa^{aa}(x,x,iu)\nonumber\\
            &&-\int_0^\infty \mathrm{d}y y^2 V_{\rm nucl}(y)\left.\left.\sum_{b,c=1}^2\left[F_\kappa^{bc}(x,y,iu)\right]^2
      \right)
\right),\nonumber
\end{eqnarray}
where $\rm Re(\dots)$ stands for the real part of the quantity in parentheses, and  $F_\kappa^{ab}(x,y,iu)$ and $G_\kappa^{ab}(x,y,iu)$
with $\{a,b\}\in \{1,2\}$ are the radial components of the free and Coulomb-dressed relativistic Green function of the virtual particles with
imaginary energy variables, respectively, and $V_{\rm nucl}$ is the radial nuclear potential.
In the case considered here, the Green function of a (bound or free) muon particle should be assumed.
To simplify the numerical evaluation, we use Eq.~(\ref{eqn:WKrho})
to calculate the potential difference $\Delta V_{\rm WK}=V_{\rm WK}^{R_{\rm nucl}\neq 0}-V_{\rm WK}^{R_{\rm nucl}=0}$ between the extended and
point-like nucleus models. Then we obtain the point-nucleus value $V_{\rm WK}^{R_{\rm nucl}=0}$ by approximate formulas from Ref.~\cite{Fai91}
adapted with the substitution $m \to m_{\mu}$. Finally, the total muonic WK potential is
\begin{equation}
V_{\rm WK}=V_{\rm WK}^{R_{\rm nucl}=0}+\Delta V_{\rm WK}\,.
\end{equation}

The electric loop VP contributions to the $g$-factor are calculated for $ns$ states by solving the radial Dirac equation numerically with the
inclusion of the above potentials, and substituting the resulting large and small radial wave function components $g_{ns}^{\mu \rm VP}(r)$ and
$f_{ns}^{\mu \rm VP}(r)$ into the integral~\cite{Bei00full}
\begin{equation}
\Delta g_{\mu \rm VP}=-\frac{8}{3}\int_0^\infty \mathrm{d}rr^3 \left(g_{ns}^{\mu \rm VP}(r) f_{ns}^{\mu \rm VP}(r) - g_{ns}(r) f_{ns}(r) \right) \,.
\end{equation}
Here, $n$ is the bound electron's principal quantum number, and $g_{ns}$ and $f_{ns}$ are the corresponding wave functions obtained without
the inclusion of the VP potentials.

\textit{Magnetic-loop vacuum-polarization contribution.} --
The muonic magnetic-loop VP contribution is calculated following the case of the corresponding VP contribution with an $e^-$--$e^+$
loop~\cite{Lee05,Yer13}. The dominant magnetic-loop contributions are described by the virtual light-by-light scattering diagrams in
Fig.~\ref{fig:fig1}~(e--f). These are obtained in the point-like nucleus assumption as~\cite{Lee05}
\begin{eqnarray}
g_{\rm ML}^{0}&=&-\frac{32}{3}\frac{\alpha(\alpha Z)^2}{\pi}\int_0^\infty \mathrm{d}qF(q m_e/m_{\mu})\\
&\times&\int_0^\infty \mathrm{d}r\left(\frac{\sin(qr)}{qr}-\cos(qr)\right)r g_{ns}(r)f_{ns}(r)\,,\nonumber
\end{eqnarray}
with the function $F(q)$ as defined and tabulated in Ref.~\cite{Lee05}. The remainder of the magnetic-loop contribution $\Delta g_{\rm ML}$
accounts for terms of higher order in $Z\alpha$, and the finite nuclear size effect for the virtual muons, and can be derived following
Ref.~\cite{Yer13} as
$\Delta g_{\rm ML}=g'_{\rm ML}-g_{\rm ML}^{0}$, where $g'_{\rm ML}$ is given by
\begin{eqnarray}
g'_{\rm ML}&=&\frac{\alpha}{\pi}\int_0^\infty \mathrm{d}\omega \mathrm{d}x \mathrm{d}y \mathrm{d}z z^3{\rm min}(x^3,y^3)\\
&\times&g_{ns}(x)f_{ns}(x)\sum_{\kappa_1,\kappa_2}\frac{4}{9}(\kappa_1+\kappa_2)^2
\left[C_1(-\kappa_1,\kappa_2)\right]^2\nonumber\\
&\times&\left[G_{\kappa_1}^{11}G_{\kappa_2}^{22}+G_{\kappa_1}^{22}G_{\kappa_2}^{11}
+G_{\kappa_1}^{12}G_{\kappa_2}^{21}+G_{\kappa_1}^{21}G_{\kappa_2}^{12}\right.\nonumber\\
&&-\left.F_{\kappa_1}^{11}F_{\kappa_2}^{22}-F_{\kappa_1}^{22}F_{\kappa_2}^{11}
-F_{\kappa_1}^{12}F_{\kappa_2}^{21}-F_{\kappa_1}^{21}F_{\kappa_2}^{12}\right]\,.\nonumber
\end{eqnarray}
Here, the angular coefficient is given in terms of a $3j$ symbol as
\begin{eqnarray}
C_1(\kappa_a,\kappa_b)&=&(-1)^{j_a+1/2}\sqrt{(2j_a+1)(2j_b+1)}\\
&\times&\left(
\begin{matrix}
j_a & 1 & j_b\\
1/2 & 0 & -1/2
\end{matrix}
\right)\frac{1+(-1)^{l_a+l_b+1}}{2},\nonumber
\end{eqnarray}
with the $j=|\kappa|-1/2$ and $l=|\kappa+1/2|-1/2$ being the total and orbital angular momentum quantum number, respectively, associated
with the relativistic quantum number $\kappa$. Finally, the total muonic magnetic-loop contribution to the $g$-factor is calculated as
$g_{\rm ML}=g_{\rm ML}^0+\Delta g_{\rm ML}$.
Both electric and magnetic loop VP contributions were checked by comparing to analytical formulas obtained to leading or all orders in
$Z\alpha$~\cite{Kar00,Lee05}, and by comparing to non-perturbative (in $Z\alpha$) calculations of $e^-$--$e^+$ VP~\cite{Bei00full,Lee05}
with the substitution $m_{\mu} \to m_e$ in our numerical computer codes.

\begin{table}[t]
\begin{center}
\caption{\label{tab:tab1}
Muonic VP contributions arising from the Uehling and WK potentials, and the magnetic loop contribution, for a range of hydrogenic
ions with charge number $Z$.}
\begin{ruledtabular}
\begin{tabular}{cccc}
    $Z$ & $\Delta g^{\rm Ue}_{\mu \rm VP}$ & $\Delta g^{\rm WK}_{\mu \rm VP}$ & $\Delta g^{\rm ML}_{\mu \rm VP}$ \\
  \hline
     1  & $-1.63\cdot 10^{-16}$     & $1.86(1)\cdot 10^{-21}$   & $1.42(1)\cdot 10^{-19}$ \\
     2  & $-2.66\cdot 10^{-15}$     & $1.19(1)\cdot 10^{-19}$   & $3.33(3)\cdot 10^{-18}$ \\
     6  & $-2.22\cdot 10^{-13}$     & $8.79(2)\cdot 10^{-17}$   & $6.17(6)\cdot 10^{-16}$ \\
    14  & $-7.13\cdot 10^{-12}$     & $1.50(1)\cdot 10^{-14}$   & $1.97(2)\cdot 10^{-14}$ \\
    18  & $-1.99\cdot 10^{-11}$     & $7.06(2)\cdot 10^{-14}$   & $4.37(4)\cdot 10^{-14}$ \\
    20  & $-3.09\cdot 10^{-11}$     & $1.36(1)\cdot 10^{-13}$   & $5.91(6)\cdot 10^{-14}$ \\
    36  & $-4.33\cdot 10^{-10}$     & $6.01(2)\cdot 10^{-12}$   & $3.26(3)\cdot 10^{-13}$ \\
    70  & $-1.48(1)\cdot 10^{-8}$   & $8.22(2)\cdot 10^{-10}$   & $5.60(6)\cdot 10^{-12}$ \\
    82  & $-4.07(1)\cdot 10^{-8}$   & $3.34(1)\cdot 10^{-9}$    & $1.17(1)\cdot 10^{-11}$ \\
    92  & $-9.44(1)\cdot 10^{-8}$   & $1.00(1)\cdot 10^{-8}$    & $2.17(2)\cdot 10^{-11}$ \\
\end{tabular}
\end{ruledtabular}
\end{center}
\end{table}

\begin{figure}[t]
   \includegraphics[width=0.98 \columnwidth]{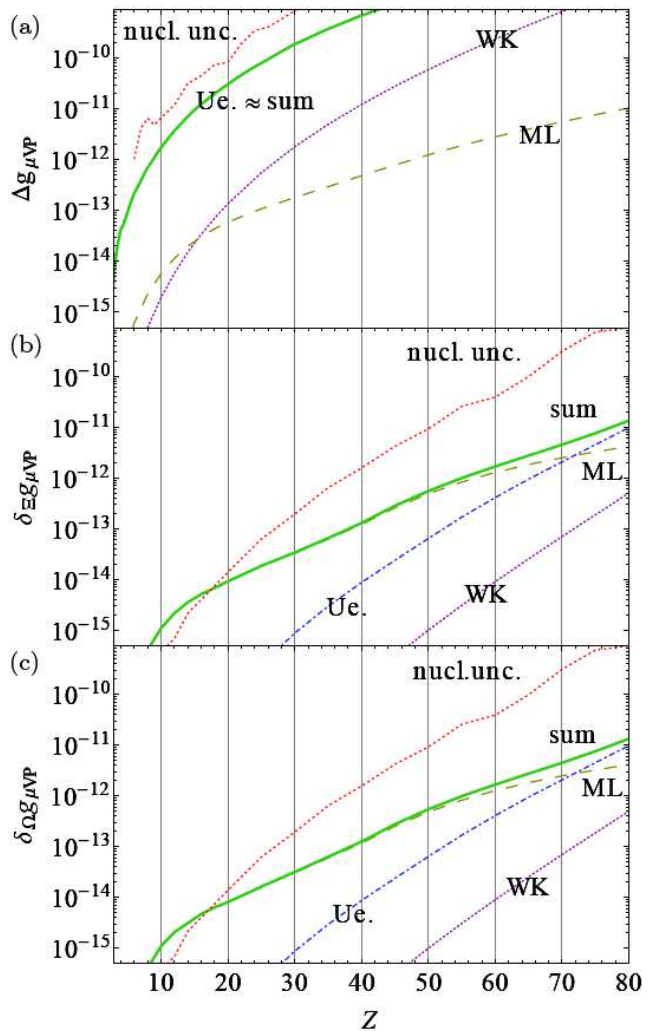}
   \caption{
   (Color online)
   (a) Muonic VP contributions arising from the Uehling (Ue, with reversed sign, dot-dashed blue) and WK (dotted violet)
   potentials, and the magnetic loop contribution (ML, dashed yellow), as well as the modulus of the total VP contribution
   (sum, solid green) for hydrogenic ions with charge number $Z$, compared to the uncertainty due to the nuclear charge distribution
   (nucl. unc., red dotted line).
   (b) The contribution of the same terms to the weighted difference of $g$-factors defined in Eq.~(\ref{eqn:weighteddiff}).
   (c) The contribution of the same terms to the second weighted difference of $g$-factors defined in Eq.~(\ref{eqn:dOg}).}
   \label{fig:fig2}
\end{figure}

\textit{Results and discussion.} --
Table~\ref{tab:tab1} contains the comparison of different electric and magnetic VP contributions to the $g$-factor of an electron bound
in the $1s$ state. The muonic VP contributions, as can be also anticipated from the scaling of the approximate, delta function-like VP
potential $-\alpha (Z\alpha) \frac{4}{15m^2} \delta(\vec{r})$ with the loop particle mass $m$, are
about $208^2 \approx 43000$ times  smaller than electronic VP in the case of the Uehling term.
Above $Z=18$, the muonic VP effects exceed the level of $10^{-11}$, which is within reach of the anticipated experimental accuracy of the
upcoming ALPHATRAP $g$-factor experiment~\cite{Stu13AP} at the Max Planck Institute for Nuclear Physics. ALPHATRAP will address specifically
intermediate and high nuclear charge states, where the muonic VP term is significant, although a direct observation is complicated
by somewhat larger uncertainties due to insufficient knowledge on nuclear charge radii or other parameters of the charge distributions,
as illustrated in Fig.~\ref{fig:fig2}~(a).
This situation is similar to the case of the muonic VP contribution to atomic binding energies~\cite{Fra91}. Recent or projected improvements of
proton distribution parameters by, e.g., collinear laser spectroscopy~\cite{Gar16}, x-ray spectroscopy of muonic atoms~\cite{AntPC} or
by electron-ion collision spectroscopy~\cite{Bra08} are anticipated to help with the experimental identification of the dominant muonic VP effect.
Furthermore, one may alternatively consider a weighted difference of $g$-factors in different charge
states with the same $Z$, in analogy with~\cite{Yer16,Sha06,Sha02}. Such a specific difference with an appropriately chosen weight factor
allows one to cancel the nuclear size dependence and the associated uncertainties by a large extent. We consider the weighted $g$-factor
difference between the $2s$ state of a Li-like ion and the $1s$ state of the H-like ion of the same $Z$,
\begin{equation}
\label{eqn:weighteddiff}
\delta_{\Xi} g=\Delta g^{(2s)}-\Xi\Delta g^{(1s)}\,.
\end{equation}
A concise formula for the weight $\Xi(Z)$ is given in Refs.~\cite{Yer16,Yer16PRA}. In order to assess the effectiveness of such a cancellation in observing
muonic effects, we approximate the muonic VP $g$-factor contribution for the Li-like ion with an effective screening potential calculated from
the probability distribution of the K-shell electrons.
We note that the same procedure may be applied to a combination of B- and H-like ions introduced in Ref.~\cite{Sha06}.

Fig.~\ref{fig:fig2}~(b) shows that while the electric loop contributions (Uehling and WK) are suppressed by 2-3 orders of magnitude with respect to
their hydrogenic values, the magnetic loop term is not largely affected by the cancellation. The reason for this is the following: the
nuclear size effect is described by a modification of the pure $-Z\alpha/r$ Coulomb potential at short distances. The radial extent of the
Uehling and WK VP potentials corresponds to the Compton wavelength of the muon of 11.7~fm, which is on the scale of nuclear radii, therefore, these
are suppressed in the weighted difference just as well as the finite nuclear size effect. However, the magnetic loop VP term describes a correction
to the coupling with the external magnetic field and thus cannot be described by a strongly localized, delta-function-like potential. Therefore, it is not
influenced much by short-scale effects such as the nuclear charge distribution.

While the absolute size of the VP contribution is significantly decreased in the weighted difference, a novel experimental concept can potentially
compensate by drastically increasing the experimental sensitivity. Taking advantage of the numerical similarity of the $g$-factors of two hydrogen- and
lithiumlike ions of different $Z$, by measuring directly the difference of the $g$-factors of two simultaneously trapped ions, experimental uncertainties
such as temporal magnetic field variations are strongly suppressed. By considering yet another weighted difference, namely, that involving two pairs
of ions with largely different  $Z$, as introduced in Refs.~\cite{Yer16,Yer16PRA}:
\begin{equation}
\label{eqn:dOg}
\delta_{\Omega} g(Z)= \delta_{\Xi}{g}(Z)-\delta_{\Xi} g([Z/2])
\end{equation}
(with $[Z/2]$ denoting the upper or lower integer part of $Z/2$), this technique can be employed to determine very small contributions that depend on $Z$,
such as the muonic VP contribution. Fig.~\ref{fig:fig2}~(c) shows that the contribution of the muonic VP terms to this difference is very similar to the
weighted difference in  Eq.~(\ref{eqn:weighteddiff}) [or see Fig.~\ref{fig:fig2}~(b)], since the muonic VP contribution to an ion with $[Z/2]$ is much
smaller than to an ion with $Z$. The combination of Eq.~(\ref{eqn:dOg}) may be rewritten as
\begin{eqnarray} \label{eq:Omega-rewritten}
\delta_{\Omega}g &=& \left(g^{(2s)}(Z)-g^{(2s)}([Z/2])\right) \\
                 &-& \Xi(Z) \left(g^{(1s)}(Z)-g^{(1s)}([Z/2]) \right) \nonumber \\
                 &-& g^{(1s)}([Z/2]) \left(\Xi(Z)-\Xi([Z/2]) \right) \nonumber \,,
\end{eqnarray}
meaning it can be efficiently determined in an experiment by measuring two $g$-factor differences, the ones in the first and second rows on the right-hand
side of the equation, and $g^{(1s)}([Z/2])$. Such a measurement with largely suppressed statistical and systematic errors is planned in the ALPHATRAP
experiment.

We note here that besides the anticipated experimental improvements, also the overall theoretical precision needs to be improved in order
to identify the muonic VP effect. It is important to note that taking advantage of the above weighted differences, a
further significant improvement of theory is not limited by nuclear effects. The evaluation of the so far uncalculated theoretical terms
is challenging but possible, and is also motivated by the prospect of an alternative determination the fine-structure constant~\cite{Yer16,Sha06}
from the bound-electron $g$-factor. In H-like ions, the most relevant missing theoretical contributions are due to two-loop QED corrections
beyond $(Z\alpha)^4$. Approx. 2/3 of the contributing Feynman diagrams have been evaluated in an all-order fashion in $Z\alpha$~\cite{Yer13PRA}.
The dominant theoretical uncertainty of $\delta_{\Xi}g$ and $\delta_{\Omega}g$ currently stems from two-body
QED effects in the Li-like ions~\cite{Yer16}. These terms can be efficiently treated by methods based on nonrelativistic QED (NRQED) expansion
theory, with a recent example reported in \cite{Puc13}. We anticipate that this Letter will further stimulate theoretical studies of the
$g$-factor of few-electron ions within the NRQED framework.

\begin{figure}[t]
   \includegraphics[width=1 \columnwidth]{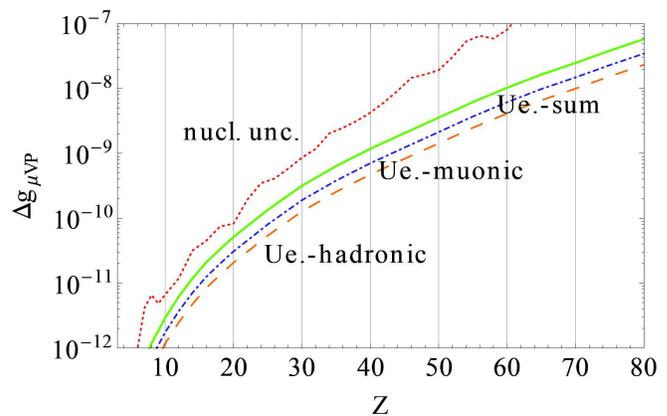}
   \caption{
   (Color online)
   Muonic (dot-dashed blue), hadronic (dashed, orange) and total (solid, green) Uehling VP contributions,
   compared to the uncertainty of the nuclear charge distribution (red dotted line).}
   \label{fig:fig3}
\end{figure}

The VP contribution due to the next heavier dilepton pair, namely, due to tauons, is more than 2 orders of magnitude weaker than the
muonic effect and therefore will not be observable for a long time, even when the weighted differences discussed in this work are
employed.
The hadronic Uehling contribution due to the virtual production of $\rho$ mesons, $\omega$ and $\psi$ vector mesons and other hadrons
can be approximated from experimental $e^{-}$--$e^{+}$ annihilation data as $0.671(15) \times \Delta g_{\mu \rm VP}^{\rm Ue}$ following
Refs.~\cite{Fri99,Eid95,Bor81}. Fig.~(\ref{fig:fig3}) shows that for several elements, the dominant (electric loop) muonic and hadronic VP effects will be
identifiable in experimental $g$-factors after an improvement of nuclear radii by approx. a factor of 5 via, e.g., the methods mentioned
above~\cite{Gar16,Bra08,AntPC}. The hadronic ML correction, as it is in leading
order described by a virtual light-by-light scattering diagram, is much more complicated to calculate. However, based on
the ratio of the hadronic and muonic light-by-light scattering contributions to the free-muon $g$-factor~\cite{Jeg09} we can estimate that
in the bound-electron $g$-factor, the inclusion of the hadronic effect further increases the magnitude of ML VP effects by 20-30\% of the
muonic contribution, and thus will be observable in future through the weighted difference (\ref{eqn:weighteddiff}) or (\ref{eqn:dOg}).

In summary, the muonic VP correction to the bound-electron $g$-factor has been evaluated. In ions with $Z>14$, due to the enhancement
of the strong Coulomb field, the magnitude of this effect can largely exceed the corresponding contribution to the free-electron $g$-factor
(5.442$\times 10^{-12}$, see Ref.~\cite{HuKi1999}). The effect is anticipated to be observable in planned trapped-ion experiments~\cite{Stu13AP,Qui01,Klu08}.
Although the muonic and hadronic VP correction is generally slightly smaller than
the uncertainty of nuclear structural contributions, these effects can be identified in future measurements after an independent improvement
of nuclear parameters or via weighted differences of the $g$-factors of H- and Li-like ions devised to cancel the detrimental nuclear effects.
The muonic VP correction will have to be taken into account for planned improvements of the fine-structure constant from the $g$-factor of highly charged ions~\cite{Sha06,Yer16}.

We acknowledge insightful conversations with Natalia S. Oreshkina. This work is part of and supported by the DFG Collaborative Research
Centre "SFB 1225 (ISOQUANT)". V.A.Y. acknowledges support by the Ministry of Education and Science of the Russian Federation (program for organizing and
carrying out scientific investigations).

\end{document}